\documentclass[superscriptaddress,aps,preprintnumbers,amsmath,showpacs,amssymb,prd,nofootinbib,preprint]{revtex4-1}
\pdfoutput=1
\usepackage{booktabs} 
\usepackage{array} 
\usepackage[table,xcdraw]{xcolor} 
\usepackage{amsmath,amssymb}
\usepackage{graphicx}
\usepackage{fancyhdr}
\usepackage{bm, color}
\usepackage{slashed,amsthm,amsfonts,empheq}
\usepackage[caption=false]{subfig}
\usepackage{hyperref}
\usepackage{booktabs}
\usepackage{multirow}
\usepackage[left=2cm,right=2cm,top=2cm,bottom=2cm,includefoot,a4paper]{geometry}

\newcommand{\Slash}[1]{{\ooalign{\hfil#1\hfil\crcr\raise.167ex\hbox{/}}}}

\newcommand{\beq}{\begin{equation}}  \newcommand{\eeq}{\end{equation}}
\newcommand{\bef}{\begin{figure}}  \newcommand{\eef}{\end{figure}}
\newcommand{\bec}{\begin{center}}  \newcommand{\eec}{\end{center}}

\newcommand{\laq}[1]{\label{eq:#1}}  
\newcommand{\Eq}[1]{Eq.(\ref{eq:#1})}

\newcommand{\eq}[1]{(\ref{eq:#1})}
\newcommand{\Sec}[1]{Sec.\ref{chap:#1}}

\newcommand{\vev}[1]{\left\langle {#1} \right\rangle}

\newcommand{\lac}[1]{\label{chap:#1}}
\newcommand{\SU}[1]{{\rm SU{#1} } }

\def\({\left(}
\def\){\right)}

\def\O{\mathcal{O}}
\def\U{\mathop{\rm U}}

\def\ebq{\end{equation} \begin{equation}}

\newcommand{\AND}{~{\rm and}~}

\newcommand{\MEV}{\,{\rm MeV}}
\newcommand{\GEV}{\,{\rm GeV}}
\newcommand{\TEV}{\,{\rm TeV}}

\def\d{\delta}
\def\e{\epsilon}
\def\f{\phi}
\def\g{\gamma}

\def\k{\kappa}
\def\l{\lambda}

\def\s{\sigma}

\def\x{\xi}
\def\y{\eta}

\def\D{\Delta}

\def\tl{\tilde}
\def\*{\dagger}

\begin{document}
%
%
%
%

\begin{flushright}
{TU-12XX}
\end{flushright}

\title{
Hubble-Scale Tachyonic Shocks from Low-Scale Inflation\\
\small{— A New Gravitational-Wave Window on Inflation}
}

\author{Haruto Masubuchi}
\affiliation{Department of Physics, Tokyo Metropolitan University, Minami-Osawa, Hachioji-shi, Tokyo 192-0397, Japan}

\author{Yuma Narita}
\affiliation{Department of Physics, Tokyo Metropolitan University, Minami-Osawa, Hachioji-shi, Tokyo 192-0397, Japan}
\affiliation{Department of Physics, Tohoku University, Sendai, Miyagi 980-8578, Japan}

\author{Wen Yin}
\affiliation{Department of Physics, Tokyo Metropolitan University, Minami-Osawa, Hachioji-shi, Tokyo 192-0397, Japan}

\begin{abstract}
Current bounds on the tensor-to-scalar ratio imply that the energy scale of inflation may lie below the grand-unified scale.
In this paper, we show that in a broad class of single-field inflation models with sufficiently small energy scales, an extremely efficient tachyonic instability develops at the end of inflation.
This instability rapidly drives the system into a nonlinear regime before coherent oscillations can be established, leading to a first-order phase-transition--like phenomenon without tunneling or barrier crossing.
The resulting ultra-relativistic shock fronts surrounding the bubble interiors expand to near the Hubble scale, corresponding to the most strongly enhanced tachyonic modes, and collide with one another, producing energetic inflaton particles and gravitational waves. As a result, the post-inflationary dynamics can differ significantly from the conventional high-scale inflationary scenario.
Interestingly, inflation at MeV--EeV energy scales can be probed via gravitational-wave observations, including pulsar timing arrays, ground-based detectors, and future space-based experiments. Recent limits from the LIGO--KAGRA--Virgo collaboration already constrain EeV-scale inflation, while pulsar timing array results may be interpreted as evidence for gravitational waves generated by GeV-scale inflation.
We also briefly discuss further implications of the resulting tachyonic shocks.
\end{abstract}
\maketitle

 \section{Introduction}

Cosmic inflation provides a compelling framework for explaining the observed homogeneity,
isotropy, and near scale-invariance of primordial density perturbations
\cite{Guth:1980zm,Sato:1980yn,Linde:1981mu,Albrecht:1982wi}.
This paradigm has been strongly supported by observations of the cosmic microwave background (CMB),
in particular through precise measurements of the amplitude and spectral tilt of the scalar power spectrum
\cite{Planck:2018vyg}.
At the same time, the absence of a detection of primordial tensor modes, characterized by the tensor-to-scalar ratio $r$,
has placed increasingly stringent constraints on high-scale inflation models.
In particular, current bounds on $r$ imply an inflationary energy scale below the grand-unified scale,
$V_0^{1/4}\lesssim 10^{16}\GEV$.

A common expectation for low-scale inflation is that observable gravitational waves sourced by
vacuum tensor fluctuations during inflation are suppressed, since the Hubble scale during inflation is small.
If the end of inflation instead involves violent, non-perturbative dynamics, additional sources of gravitational waves can arise
\cite{Kofman:1994rk,Kofman:1997yn,Felder:2000hj,Felder:2001kt,Khlebnikov:1997di,Garcia-Bellido:2007fiu}.
In these studies, coherent oscillations of the inflaton lead to parametric or tachyonic resonance,
through which the inflaton condensate is fragmented by nonlinear dynamics.
The resulting highly inhomogeneous field configurations act as efficient sources of gravitational waves.
However, the characteristic frequencies of these gravitational waves are typically high, often higher than the GHz range.
 This is because the relevant dynamics is controlled by the oscillation timescale of the inflaton, which is much shorter than a Hubble time. 
Such high-frequency gravitational waves are beyond the reach of current gravitational-wave detection technologies.
In order to generate gravitational waves at observable low frequencies,
the nonlinear dynamics may instead be governed by around the Hubble timescale,
as is the case for bubble-wall dynamics in a first-order phase transition~\cite{Witten:1984rs,Hogan:1986qda,Caprini:2015zlo}.

In this work, we show that such Hubble-scale non-perturbative dynamics is expected in a wide class of single-field inflation models once the inflationary energy scale is sufficiently low.
For sufficiently low-scale inflation, the curvature of the potential near the end of inflation necessarily changes significantly in order to satisfy the CMB normalization. As a result, the end of inflation is naturally accompanied by a pronounced tachyonic instability, which leads to the exponential growth of horizon-scale inflaton fluctuations even before the onset of coherent oscillations.
While this effect is subdominant in high-scale inflation, it becomes inevitable at sufficiently low scales once the scalar perturbation amplitude is fixed to its observed value.

As a result, inflaton fluctuations rapidly enter the nonlinear regime,
locally reaching large field amplitudes.
These nonlinear fluctuations provide seeds for a strongly first-order phase transition-like phenomenon,
leading to the formation of shock waves and bubble-like structures expanding at relativistic speeds.
The typical separation between these bubbles is of order the Hubble scale,
reflecting the fact that the tachyonic instability is triggered at the end of inflation.
Similar to a first-order phase transition, this process provides an efficient source of gravitational waves,
with characteristic frequencies set by the inflationary Hubble scale.
This phenomenon therefore offers a new observational window into low-scale inflation.

Bubble collisions in the context of tachyonic preheating were discussed in the original works
\cite{Felder:2000hj,Felder:2001kt}
(see also \cite{Antusch:2015vna,Amin:2018kkg, Kitajima:2018zco} for related phenomena).
In these studies, tachyonic instability plays an important role in amplifying field fluctuations and inducing
inhomogeneous structures.
However, the tachyonic instability does not itself trigger the bubble dynamics in the sense relevant for
first-order phase transition--like phenomena.
Rather, it mainly leads to the rapid growth and decoherence of field fluctuations inside already expanding regions. 

In hybrid inflation, tachyonic preheating and the associated formation of bubbles or topological defects occur due to the dynamics of the waterfall field \cite{Felder:2000hj,Felder:2001kt,Dufaux:2008dn}, which is initially trapped at a symmetry-enhanced point (see, however, e.g., \cite{Narita:2023naj}, where the mechanism discussed in this work may also apply%
\footnote{Although the criterion \Eq{nonlinear} does not directly apply, the waterfall field in \cite{Narita:2023naj}, which has a potential of the form $V_0 - \l \f^{4}$, can induce similar dynamics. This is because, in that model, the inflaton and waterfall fields are smoothly connected without branching, and there exists a parameter region in which the waterfall field effectively drives the end of inflation.}).
As a consequence, the characteristic length and time scales of the resulting nonlinear dynamics—and hence the peak frequency, the typical bubble size at collision, and the amplitude of the generated gravitational waves—are controlled by the mass scale of the waterfall field, which governs the tachyonic instability. Bubble formation of the Higgs field, driven toward the wrong vacuum away from the electroweak one due to enhanced fluctuations induced by inflaton tachyonic resonance, has also been discussed in Ref.~\cite{Li:2022ugn}. In contrast, we consider single-field inflation, where the inflaton is not initially located at a symmetry-enhanced point.
In this case, the nonlinear dynamics is triggered directly by the inflaton itself,
and the characteristic scale of the bubble-like structures, as well as that of the resulting gravitational-wave signal,
is set by the inflationary Hubble parameter, which is the instability scale around the end of inflation.

\section{Setups}\lac{2}

\paragraph{Equations}
We decompose the inflaton field $\f$ as
\beq
\f=\bar{\f}+\d\f,
\eeq
where $\bar \f$ denotes the spatially averaged (homogeneous) mode and $\d\f$ the fluctuation.
The classical equation of motion is
\beq
\ddot{\f} +\frac{\nabla^2}{a^2}\f +3 H \dot \f = -\partial_\f V,
\eeq
where $\bar \f$ {and $\d\f$} denotes the spatially averaged (homogeneous) mode and
the fluctuation{, respectively}, $a$ is the scale factor and $H=\dot a/a$ is the Hubble parameter.
We focus on the era when the energy density is dominated by the inflaton sector and $H$ follows the Friedmann equation{:
\begin{equation}
    \label{eq:Fried}
    H^2 = \frac{\rho_\phi}{3 M_\text{pl}^2}, \quad
    \rho_\phi = \frac{1}{2} \dot{\phi}^2 + \frac{1}{2 a^2} (\nabla \phi)^2 + V(\phi).
\end{equation}
Here $M_{\rm pl}=2.4\times 10^{18}\GEV$ is the reduced Planck mass.
}

We first review the inflationary era, which is largely governed by the homogeneous mode $\bar \f$.
We then discuss the tachyonic instability shortly after the end of inflation by focusing on the evolution of $\d\f$ in the next section.

\paragraph{ Low-scale inflation models}
We assume that the inflaton potential around the slow-roll region takes the form
\beq
V {\approx} V_0-\l \f^{2n} 
\laq{approx}
\eeq
with $\l>0$. See \Sec{discussions} for the generalization to potentials with more terms. 
This type of low energy inflation model can arise from UV models that stabilize the radiative corrections such as by considering the axionic realization~\cite{Czerny:2014wza,Czerny:2014xja,Daido:2017wwb,Takahashi:2019qmh}, non-minimal coupling to gravity~\cite{Yin:2022fgo}, and supersymmetry~\cite{Takahashi:2013cxa, Murai:2023gkv}, and heavy QCD axion inflation~\cite{Takahashi:2021tff}. 
Note that when we consider an inflation at a large field value, still one can obtain the similar potential by expanding around the field for the inflation to happen, e.g.,~\cite{Yin:2022fgo}.
We use the notation that for $n=2,3,\cdots$ the model corresponds to hilltop inflation.\footnote{
It is well known that the cases $n=1/2$ and $n=1$ are not successful for low-scale inflation, because
neglecting $\e$ the slow-roll parameter $\eta$ is either too small or does not change in time.
This implies that higher-order terms in the potential become important, and again the condition for the tachyonic shock holds (see \Sec{discussions}). }
For half-integer values $n=3/2,5/2,\cdots$, the model instead corresponds to inflection-point inflation.
In particular, $n=2$ realizes quartic hilltop inflation.
Without loss of generality, we consider slow-roll inflation to proceed in the direction of increasing $\f$.

\paragraph{Conditions for inflation}
Slow-roll inflation occurs when the slow-roll parameters
\beq 
\e \equiv \frac{M_{\rm pl}^2}{2} \left(\frac{V'}{V}\right)^2,
\qquad
\y \equiv M_{\rm pl}^2 \frac{V''}{V},
\qquad
\x^2 \equiv M_{\rm pl}^4 \frac{V'}{V} \frac{V'''}{V}
\eeq
are sufficiently small, where $X'[y]\equiv dX/dy$.
The scalar power spectrum of curvature perturbations is given by
\beq\laq{amplitude}
\Delta_{\cal R}^{2}(k)
= \frac{1}{24\pi^{2}}\frac{V}{M_{\rm pl}^4\epsilon}.
\eeq

In low-scale inflation, $\e$ must be extremely small in order to reproduce the observed amplitude of the power spectrum.
First, this implies that during inflation $V\simeq V_0$, and the inflationary Hubble scale is
$H_{\rm inf}\simeq \sqrt{V_0/(3M_{\rm pl}^2)}$.
Second, the end of inflation is determined by the second slow-roll parameter $\eta$,
$
-\eta \approx 1$.
Very soon after this condition is satisfied, $\e$ rapidly grows to unity and the accelerated expansion terminates.

Finally, neglecting subdominant contributions from $\e$, the observables can be written as
\beq 
n_s \simeq 1+2\y,
\qquad
r \simeq 16\e,
\qquad
\frac{dn_s}{d\ln k} \simeq -2\x^2.
\eeq

{
Under {the potential \eq{approx} and the slow-roll approximation},
the spectral index, the tensor-to-scalar ratio, and the running of the spectral index are
{respectively derived as}
\beq
\laq{ns} n_s
\simeq
1-\frac{2n-1}{(n-1)N},
\ebq
r
\simeq
\frac{2V_0}{3\pi^2 A_S M_{\rm pl}^4},
\ebq
\frac{dn_s}{d\ln k}
\simeq
-\frac{2n-1}{(n-1)N^2}
\laq{run}
\eeq
where $A_S=\D_{\cal R}^2(k_*)$ is the amplitude of the scalar power spectrum evaluated at the pivot scale $k_*=0.05\rm Mpc^{-1}$.
In deriving these expressions, we neglect terms suppressed by $\f$ at the end of inflation.
Here, $N$ denotes the number of $e$-folds between the horizon exit of the pivot scale and the end of inflation.
}
Assuming instantaneous reheating, the number of $e$-folds can be estimated from the expansion history
from the end of inflation until the reentry of the CMB pivot scale,
\beq
N \approx 28 + \ln\!\left(\frac{V_0^{1/4}}{10\TEV}\right) \laq{efold}
\eeq

From the observational side, the measured amplitude of the curvature power spectrum at the CMB scale is
\cite{Planck:2018jri}
\beq
A_{S,{\rm obs}} \approx 2 \times 10^{-9}.
\eeq
The spectral index measured by {\it Planck} is \cite{Planck:2018jri}
$
n_{s,{\rm obs}} = 0.965 \pm 0.004,
$
where we have adopted the result of the {\it Planck} TT, TE, EE + lowE analysis.
On the other hand, the ACT collaboration~\cite{ACT:2025fju,ACT:2025tim},
combined with CMB measurements from BICEP/Keck~\cite{BICEP:2021xfz} and {\it Planck}
\cite{Planck:2018jri,Planck:2018vyg}, as well as baryon acoustic oscillation data from DESI
\cite{DESI:2024mwx}, reports a slightly higher value of the spectral index,
$
n^{\rm ACT}_{s,{\rm obs}} = 0.974 \pm 0.003.
$
The running of the spectral index is constrained to be
$
\frac{dn_{s,{\rm obs}}}{d\log k} = -0.0045 \pm 0.0067,
$
where we again adopt the result of the {\it Planck} TT, TE, EE + lowE analysis \cite{Planck:2018jri}.

Given this small value of $N$, the observed spectral index $n_{s,\rm obs}$ is difficult to explain
within low-scale inflation models with $N<50$.
However, in low-scale inflation the spectral index is highly sensitive to small modifications of the potential.
For example, a tiny linear term~\cite{Takahashi:2013cxa},
a Coleman--Weinberg correction, or lower-dimensional operator corrections~\cite{Daido:2017tbr}
can significantly improve the fit.
Such modifications do not change the amplitude of the power spectrum very significantly.

We also note that the running of the spectral index is enhanced when $N$ is small,
which serves as a generic prediction of low-scale inflation.
This can be probed at the level of
$\frac{dn_s}{d\ln k_*} \gtrsim 10^{-3}$,
for example by SPHEREx~\cite{Dore:2014cca},
and in combination with DESI~\cite{DESI:2013agm}, WFIRST~\cite{spergel2015wide},
or SKA~\cite{CosmologySWG:2015ysq},
leading to \Eq{run} with $N \lesssim 30\text{--}40$ within reach.
Intuitively, the smaller the required value of $N$, the more rapidly the curvature of the potential
must change in order to satisfy the condition for the end of inflation and the measured the spectral index.
Indeed, the running is further enhanced when mechanisms are introduced to increase the spectral index from \Eq{ns}
around horizon exit,
see, e.g.,~\cite{Daido:2017tbr,Takahashi:2019qmh,Yin:2022fgo}.

\paragraph{Relevant relations for preheating era}
In the following, we adopt the amplitude predicted in typical low-scale inflation models.
For $n>1$, imposing the requirement that the observed amplitude of the power spectrum
(i.e., the CMB normalization) is reproduced, we obtain
\beq
\lambda V_0^{n-2}
\simeq
\frac{\left(48\pi^2 n^2 A_S\right)^{n-1} M_{\rm pl}^{2(n-2)}}
{\left[4n(n-1)N\right]^{2n-1}}.
\laq{lambda}
\eeq
The end of inflation is estimated as 
\beq 
-\eta\approx 1\;\;\leftrightarrow\;\;
3H_{\rm inf}^2 \approx -V''(\bar\f_{\rm inf})
=2n(2n-1)\l \bar\f_{\rm inf}^{2n-2}
\equiv m_{\rm eff}^2(\bar\f_{\rm inf}),
\laq{endinf}
\eeq
where $m_{\rm eff}^2(\overline{\phi})$ denotes the (tachyonic) effective mass squared. 
Here, the time at which inflation ends, $t=t_{\rm inf}$, is defined by
\beq
\bar\f_{\rm inf} \equiv \bar\f(t_{\rm inf}).
\eeq
Then, the background field value at the end of inflation is estimated as
\beq
\overline{\phi}_\text{inf}
\approx
\frac{H_{\rm inf}}{\sqrt{A_S}\,\pi}\,
\bigl((n-1)N\bigr)^{\frac{2n-1}{2n-2}}\,
\(\frac{2}{2n-1}\)^{\frac{1}{2n-2}}.
\eeq

\section{Tachyonic shocks as preheating in low-scale inflation}

The explosive preheating proceeds through three distinct stages, which we describe in turn.
We present a generic, semi-analytical discussion of Stage 1. For simplicity, in stage 2 and 3 we assume that the potential is stabilized by higher-order terms and develops a minimum around
$v = (V_0/\lambda)^{1/(2n)}$,
with the corresponding vacuum mass of order $\sqrt{V_0/v^2}$.

In parallel, we confirm each process using two-dimensional lattice simulations, in which we consider the following two benchmark potentials with $n=3/2$ and $n=2$, respectively:
\beq
 V\simeq V_0\(\frac{27}{256} -\(\frac{\phi}{v}\)^{3}+\(\frac{\phi}{v}\)^4\), \AND~~
 V_0\(\frac{4}{27}-\(\frac{\phi}{v}\)^{4}+\(\frac{\phi}{v}\)^6\).
 \laq{benchmarks}
\eeq
In the simulations, we enhance the initial fluctuations by hand,
due to the prohibitive requirement of an extremely large number of lattice points
needed to fully capture the dynamics in the parameter region of interest.
We consider this procedure to be sufficient for demonstrating the relevant phenomena
in the subsequent two stages, since the enhancement of fluctuations in the linear regime
during Stage~1 is well understood as tachyonic instability
\cite{Felder:2000hj,Felder:2001kt}.
The detailed setup of the numerical simulations is provided in the Appendix \ref{app:1}.

\subsection{Stage 1: Explosive Tachyonic Instability}
For arbitrary $n>1$, after inflation ends, the magnitude of the curvature $m_{\rm eff}^2=|V''(\f)|$ subsequently increases from $H_{\rm inf}$ satisying \Eq{endinf} as the field rolls away from the origin $\f=0$.
This stage is short compared to a Hubble time, since the typical timescale is set by $m_{\rm eff}$,
which becomes larger than $H_{\rm inf}$ shortly after the end of inflation.

We can study the first stage of the evolution of the system
within the linear approximation of $\d\f$,
\beq\laq{linear}
 \ddot{\bar\f}=-\partial_\f V(\bar \f),~\AND~~
 \d \ddot{\f}_k = -\vec{k}^2 \d \f_k + m_{\rm eff}^2[\bar\f] \d \f_k .
\eeq
Here $k\equiv |\vec{k}|$ is the magnitude of the physical momentum $\vec{k}$.
The first equation admits the approximate solution
\beq
\bar\phi \simeq 
\left[(n-1)\sqrt{2\lambda}\,(t-t_*)\right]^{\frac{1}{1-n}},
\eeq
where $t\to t_*$ corresponds to the rapid increase of $\bar\phi$.
From this expression, one can see that the characteristic timescale of the evolution is set by $t-t_*\sim m_{\rm eff}^{-1}$.
From the second equation, 
one immediately finds tachyonic growth when $k\ll m_{\rm eff}$. 

{Neglecting the gradient term,}
and substituting the background solution $\bar\f$ obtained from the equation of motion,
one can derive a power-law growth~\cite{Felder:2000hj,Felder:2001kt}
\beq
\d\f_k
\simeq
\d\f_{k,\rm inf}
\left(\frac{\bar\f}{\bar\f_{\rm inf}}\right)^{n},
\eeq
which is valid until the breakdown of the linear approximation.
This scaling can be confirmed from panel~\textcircled{1} in Figs.~\ref{fig:lattice} and \ref{fig:lattice2}
at the early stage, where we also plot $\s_\f/\f^{n}$ by the red dashed lines, with $\s_\f$ denoting the measured variance of the field.
One finds that for $n>1$, the fluctuation grows faster than the background field $\bar\f$.

Furthermore, for $\bar\f[t]>\bar\f[t_e]$, a 
Minkowski fluctuation that enters the tachyonic regime at $t_e$
grows as
\beq
\laq{scaling}\frac{k^3}{2\pi^2}\vev{|\d \f_{k}[t]|^2}
=
\left(\frac{\bar\f[t]}{\bar\f[t_e]}\right)^{2n}
\frac{k^3}{2\pi^2}\vev{|\d \f_{k}[t_e]|^2}.
\eeq
The Minkowski fluctuation just before entering the tachyonic regime 
has a typical amplitude 
$\delta \phi_k \sim 1/\sqrt{k}$, 
where we have neglected the mass term in the regime $k \gg m_{\rm eff}$. 
The mode enters the tachyonic regime when 
$k \sim m_{\rm eff}(t_e)$, 
as $m_{\rm eff}$ grows with time, which determines $\bar{\phi}(t_e)$. 
Using the scaling $\delta \phi_k \propto \bar{\phi}^n$ in the tachyonic phase 
together with $m_{\rm eff} \propto \bar{\phi}^{\,n-1}$, we obtain
\beq
\delta \phi_k(t)
\simeq
\frac{1}{\sqrt{k}}
\left(\frac{m_{\rm eff}(t)}{k}\right)^{\frac{n}{n-1}}
\Theta\!\left(m_{\rm eff}(t)-k\right).
\eeq
When $k\sim m_{\rm eff}[t]$ the mode is as large as the Minkovski one, but the smaller the momentum, the larger enhanced fluctuation~\cite{Felder:2001kt}. Indeed, this fluctuation is always more IR dominant compared to the scale-invariant fluctuation for $n>1$. 
In our setup, the most IR mode is cutoff by the Hubble parameter, and enters into the tachyonic regime at the end of inflation, which is the reason we set the IR cutoff by using the Heaviside step function. 
In the following, we therefore take $\d\phi[t_1]= (\bar{\phi}/\bar \phi_{\rm inf})^n \d \f_{\rm inf}$ suppress the explicit momentum dependence, where we take $\d \f_{\rm inf}=H_{\rm inf}/(2\pi).$

\begin{figure}[!t] 
    \begin{center}
        \includegraphics[width=150mm]{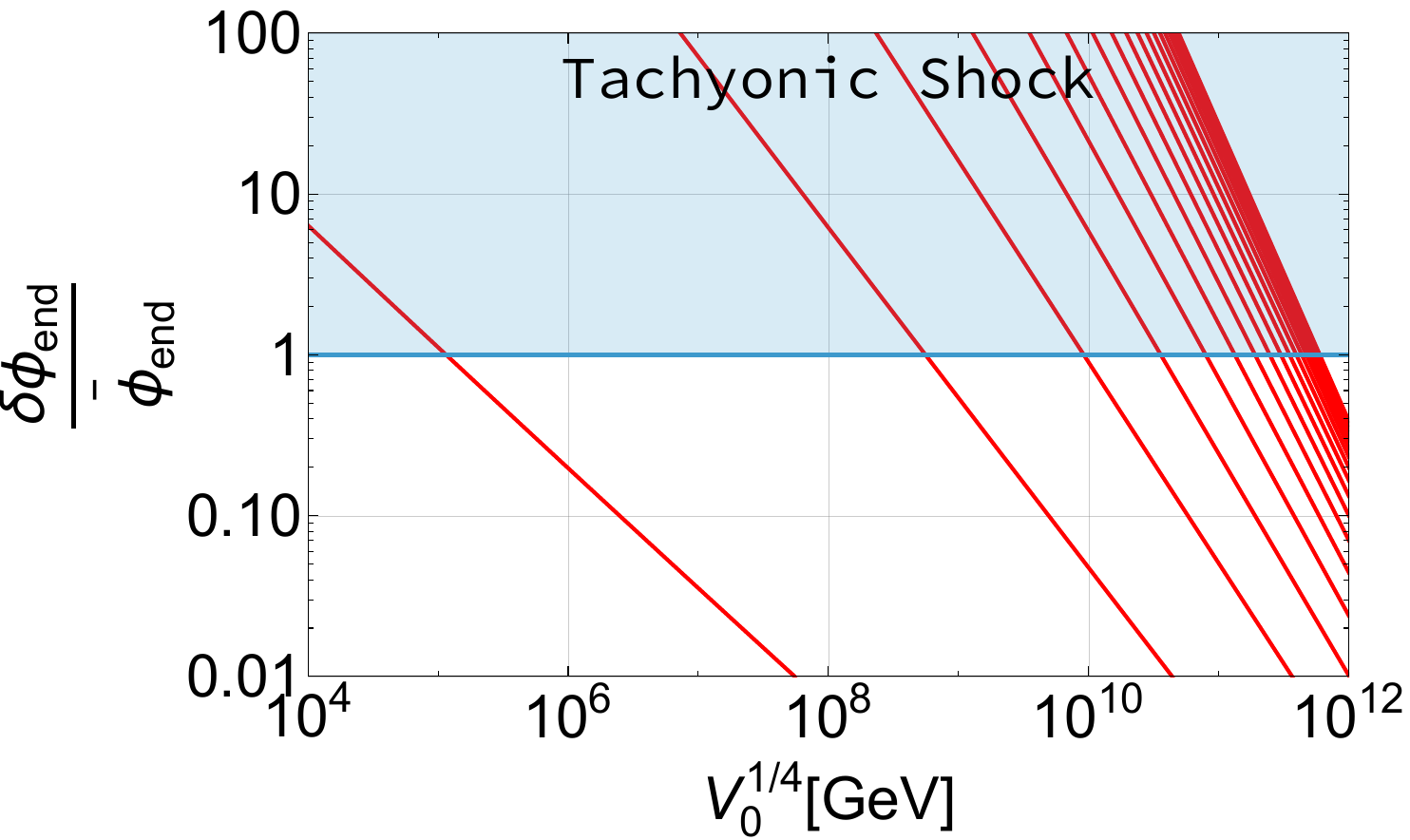}
\vspace{-5mm}
    \end{center} 
\caption{
$\d\phi_{\rm end}/\bar\phi_{\rm end}$ with $\k=1$ as a function of the inflationary energy scale $V_0^{1/4}$.
The red curves correspond to $n=3/2,2,\ldots,10$ from left to right.
In the blue shaded region, the linear perturbation is violated during the rolling of $\bar\phi$ toward $\bar\phi_{\rm end}$, and a tachyonic shock occurs. \Eq{efold} and $A_S=2\times 10^{-9}$ are used. 
}
    \label{fig:1} 
\end{figure}

For simplicity, let us assume that higher-order terms become important when the potential energy
changes by $\O(V_0)$, i.e., these terms prevent the potential from entering a negative-energy regime.
Accordingly, we parametrize the endpoint of our analysis as
\beq
\bar \f_{\rm end}= \k v, \AND
v \equiv \(V_0/\l\)^{\frac{1}{2n}}.
\eeq
 $\k$ is an $\O(1)$ model-dependent parameter.
Then, we obtain
\begin{align}
    \frac{\d \f_{\rm end}}{\bar \f_{\rm end}} &\approx
    2^{-\frac{1}{n}-\frac{n}{2(n-1)}}
    [(n-1)N]^{-\frac{(2 n-1)^2}{2 (n-1) n}}
    (2n-1)^{\frac{n}{2(n-1)}}
    \kappa^{n-1}
    \sqrt{A_S}
    \left(\frac{H_\text{inf}}{\sqrt{n A_S} \pi M_\text{pl}}\right)^{\frac{1-n}{n}} 
      \laq{generic}
\end{align}
The exponent of $H_{\rm inf}$ is negative, implying that the smaller the inflation scale,
the larger the relative fluctuation amplitude. Figure~\ref{fig:1} shows the result for various $n$ by varying $V_0^{1/4}$.  
{In particular, for any $n>1$, if $H_{\rm inf}$ is sufficiently small, the condition
$
\d\f \sim \bar\f
$
is reached before higher-order terms become important. This region is shaded in blue in Fig.\ref{fig:1}. 
As we will see, an explosive preheating phenomenon occurs,
analogous to a first-order phase transition triggered by the onset of nonlinearity,
which we refer to as a {\it tachyonic shock}.
One finds
\beq
\boxed{\text{tachyonic shock if } 
V_0^{1/4}\k^{-n/2} \lesssim 
10^{5}\GEV \ (\text{inflection}) \AND 
10^{9}\GEV \ (\text{hilltop})}
\laq{nonlinear}
\eeq
for $n=3/2$ and $n=2$. As seen in the figure, for any finite $n>1$,
the threshold lies below $\sim 10^{12}\,\mathrm{GeV}$,
and is minimized for $n=3/2$ and $n=2$.\footnote{If a finite reheating period is taken into account,
the effective $N$ becomes smaller and the tachyonic shock is easier to occur.}
Therefore, once the condition \Eq{nonlinear} is satisfied,
the tachyonic shock develops.
}

This is the regime of interest in this work and has not been discussed extensively in the literature.
$\bar\f\approx \d \f$ occurs when
\beq
\k_{\rm eff}\equiv \frac{m_{\rm eff}}{H_{\rm inf}}\bigg|_{\text{end of Stage~1}}
\approx
\sqrt{6} (n-1) N
\left(\frac{(n-1)N}{2 n-1}\right)^{\frac{1}{2 (n-1)}}
A^{-1/2}_S
\laq{timescale}
\eeq
which is, interestingly, independent of the inflationary energy scale. $\k_{\rm eff}\approx 10^{6-7}$ from the CMB normalization, $n=3/2$--$10$ and $N=20$--$50$.

\subsection{Stage 2: Nonlinear shock wave and bubble expansion}

\begin{figure}[!t] 
    \begin{center}
        \includegraphics[width=170mm]{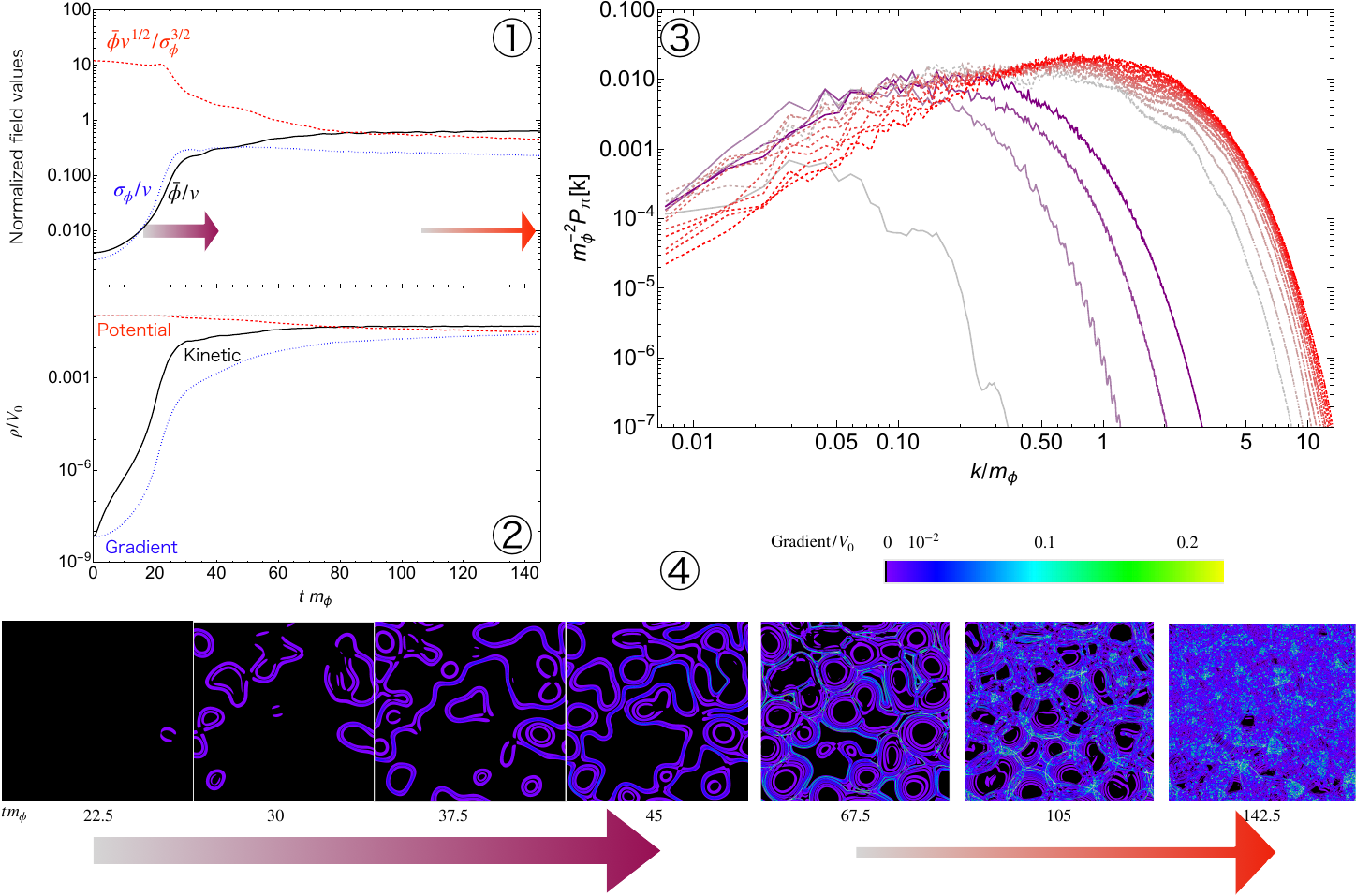}
\vspace{-20mm}
    \end{center} 
\caption{
Two-dimensional lattice simulation for the case $n=3/2$, corresponding to an inflection-point inflation model.
The upper-left panels labeled \textcircled{1} and \textcircled{2} show the time evolution of the scalar field
and the energy components, respectively.
The broad purple arrow and the narrow red arrow indicate the benchmark time evolution at which we present
the reduced power spectra of $\dot{\phi}$ and snapshots of the gradient energy, labeled \textcircled{3} and \textcircled{4}, respectively.
In panel \textcircled{3}, the purple solid lines correspond to the broad purple arrow,
while the red dashed lines correspond to the narrow red arrow.
See Appendix~\ref{app:1} for details of the numerical simulations.
Here $m_\phi = (3/2)\sqrt{V_0}/v$ denotes the vacuum mass.
}
    \label{fig:lattice} 
\end{figure}

\begin{figure}[!t] 
    \begin{center}
        \includegraphics[width=170mm]{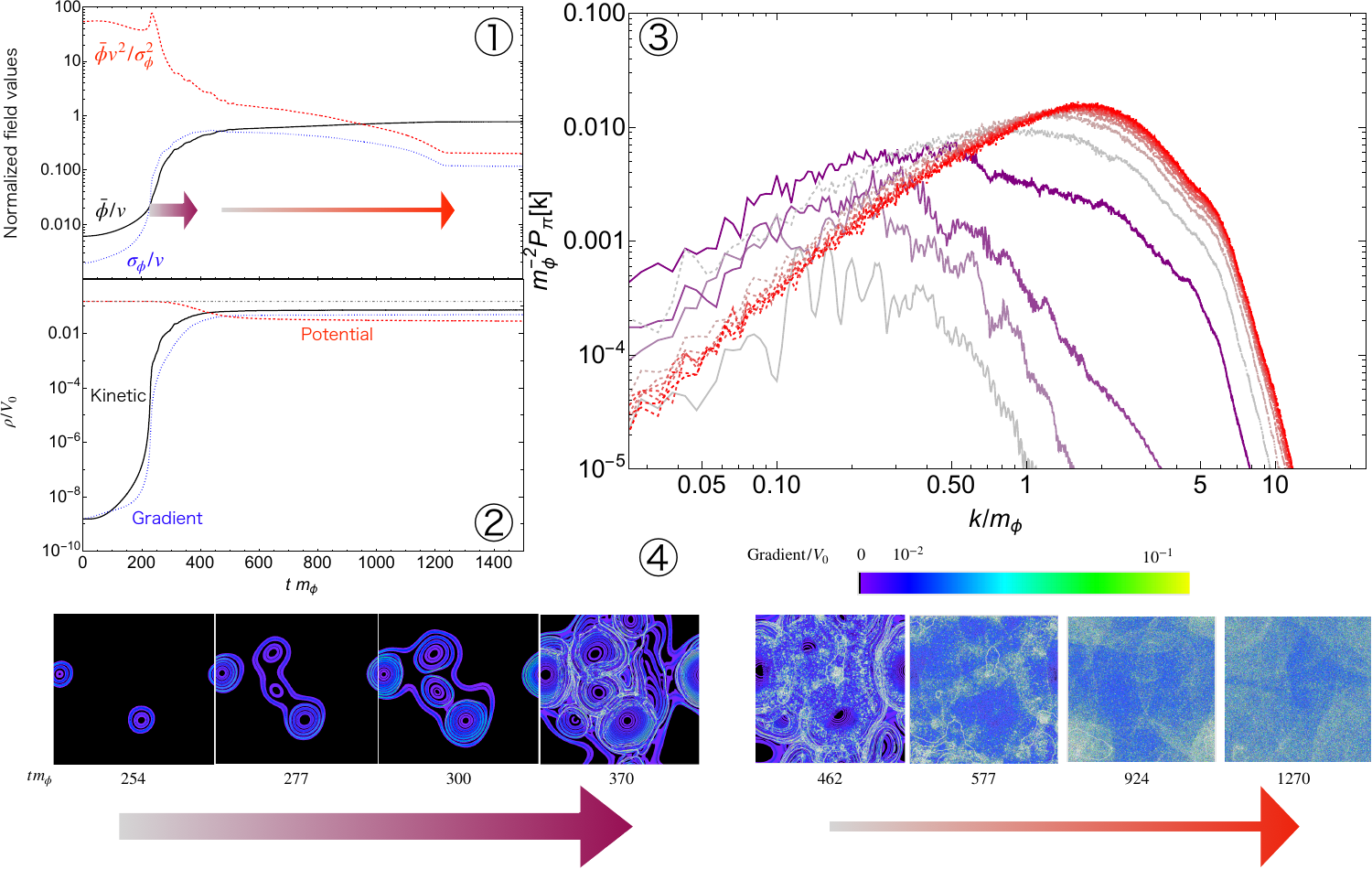}
\vspace{-20mm}
    \end{center} 
\caption{
Same as \ref{fig:lattice} but with $n=2$, i.e., a hilltop inflation. 
Here $m_\phi =4/\sqrt{3} \sqrt{V_0}/v$ is the vacuum mass.
}
    \label{fig:lattice2} 
\end{figure}

When $\delta\phi$ becomes comparable to $\bar\phi$, the linearized equation \Eq{linear} is no longer applicable.
This does not terminate the growth of fluctuations; rather, it further enhances it, since in the nonlinear regime with $|\delta\phi|\gg\bar\phi$, the potential term in the equation of motion behaves approximately as
$\propto |\delta\phi|^{2n-2}\,\delta\phi$.
As a result, regions with larger field amplitudes evolve more rapidly
(see the sudden jump of $\bar\phi$ and $\sigma_\phi$ when they cross each other in panel~\textcircled{1} of Figs.~\ref{fig:lattice} and \ref{fig:lattice2}).

Some localized regions of the field evolve significantly earlier than their surroundings;
we refer to these regions as \emph{seeds}. These earliest seeds subsequently develop into expanding bubble-like configurations.
In contrast, the field remains smaller in most other regions.
The typical separation between seeds is of order $1/H_{\rm inf}$, 
corresponding to the momentum scale of the tachyonically enhanced modes 
as determined by the condition \eq{endinf}.

A bubble can expand if the vacuum energy released inside 
overcomes the gradient energy associated with the wall.
More precisely, the initial field configuration must extend over a spatial region 
of size $\D L>c_L/m_{\rm eff}$  in order for the local evolution to proceed efficiently,
since the rolling timescale is $c_L/m_{\rm eff}$ while gradient effects propagate at most at the speed of light.\footnote{This condition is verified using the one-dimensional analysis based on \Eq{eq1d}.} Here $c_L\gtrsim \O(1)$ parametrizes the nonlinear correction to the rolling timescale. 
In particular, the typical length scale of the enhanced modes is of order $1/H_{\rm inf}$,
which is much larger than $1/m_{\rm eff}$.
Therefore, the seeded configurations naturally evolve into expanding bubbles, with the typical size of the bubble increase in time.

As the outer gradient wave propagates outward, 
its thickness decreases due to Lorentz contraction. 
For an expansion over a scale $R$, 
the released vacuum (latent) energy is $\Delta E\sim V_0 R^3$. 
This energy is stored in the kinetic and gradient energy of the scalar wave front 
and rapidly becomes the dominant component of the energy density. 
Estimating the wall tension as $\sigma\sim V_0^{1/2} v$, 
energy conservation implies a Lorentz factor $\g_w\sim V_0^{1/2}/v \times R$. When the condition \Eq{nonlinear} is satisfied, 
this leads to the formation of expanding shock-wave--like structures 
with $\gamma_w\gg 1$, 
analogous to relativistic bubble walls in a first-order phase transition, 
but without tunneling or barrier crossing.\footnote{Bubble formation without potential barrier has been discussed in the literature
\cite{Lee:1985uv,Felder:2001kt} in the context of quantum tunneling and with multi-fields~\cite{Li:2022ugn}.}
Oscillatory wave structures may develop inside the bubbles (see snapshot~\textcircled{4} of Fig.~\ref{fig:lattice2}); however, the dominant fraction of the energy is carried by the outermost propagating wave front. The dominance of the gradient energy can be seen in the snapshots of the gradient energy
labeled \textcircled{4}, as well as in the time evolution of the energy components
labeled \textcircled{2} in Figs.~\ref{fig:lattice} and \ref{fig:lattice2}.
The Lorentz contraction of the wall width is also clearly visible in these snapshots.

Strictly speaking, after the formation of the bubble, 
at a time $\Delta t_{\rm osc}=1/(\kappa_{\rm eff}H_{\rm inf})$, 
the outer region begins to evolve and develops inhomogeneities. 
A useful diagnostic is the gradient energy density (or the spatially averaged squared gradient) outside the bubble,
\beq
\vev{(\nabla\phi)^2}_{ob}\ \propto\ \int d\log k\, k^{d}\,k^2\,|\delta\phi_k|^2
\ \propto\ \int d\log k\, k^{d-\frac{n+1}{n-1}},
\eeq
where $d$ is the spatial dimension and we used the linear-theory estimate
$|\delta\phi_k|\propto k^{-1/2-n/(n-1)}$,
which remains valid when we restrict attention to the spatial region outside the bubbles,
where the field fluctuations are still in the linear regime.
For $n\gtrsim (d+1)/(d-1)$, the integral is UV dominated, indicating that short-wavelength fluctuations
can contribute significantly to $\vev{(\nabla\phi)^2}$ in the outer region.
However, modes near the UV end are only marginally tachyonic and therefore are not efficiently amplified.
Consequently, such short-wavelength fluctuations typically do not develop into many independent expanding bubbles in the outer region.
The seeded bubble continues to expand.
An animation from the two-dimensional lattice simulation for the gradient energy for $n=5$ 
is provided in the supplementary material (\texttt{n5.gif}, see Appendix~\ref{app:1}).

In the complementary case $n\lesssim (d+1)/(d-1)$, the integral is IR dominated.
In this situation, short-wavelength fluctuations cannot compete with the large-scale gradients induced by the dominant long-wavelength modes, except possibly near their local extrema.
Around the extrema of the Hubble-scale modes, fluctuations—particularly those with relatively long wavelengths—can undergo shock-like evolution similar to that of the primary seeds.
Nevertheless, these modes have only a few extrema per Hubble volume, with a typical separation of order $1/H_{\rm inf}$.
Instead, bubbles tend to form in localized clusters around these extrema and subsequently expand
(see panel~\textcircled{4} in Fig.~\ref{fig:lattice2}).
As a result, away from those extrema the intermediate region mainly oscillates with inhomogeneities rather than forming additional bubbles.

Since the width of the relativistic wave front is
$L_w \sim \gamma_w^{-1}\sqrt{v^2/V_0},
$
which is much shorter than the oscillation timescale, the shock wave does not stall due to the field evolution in the outer region. 
This is because the oscillating scalar field spends most of its time, 
$\Delta t_{\rm osc} \, (\gg \sqrt{v^2/V_0})$, near its original position. As a result, the bubbles primarily interact with the field around the origin 
and continue extracting the available latent energy.
An animation of the two-dimensional lattice simulation for the case $n=3/2$
is provided as supplementary material (file \texttt{n32.gif}; see Appendix~\ref{app:1}).

\begin{figure}[!t] 
    \begin{center}
        \includegraphics[width=150mm]{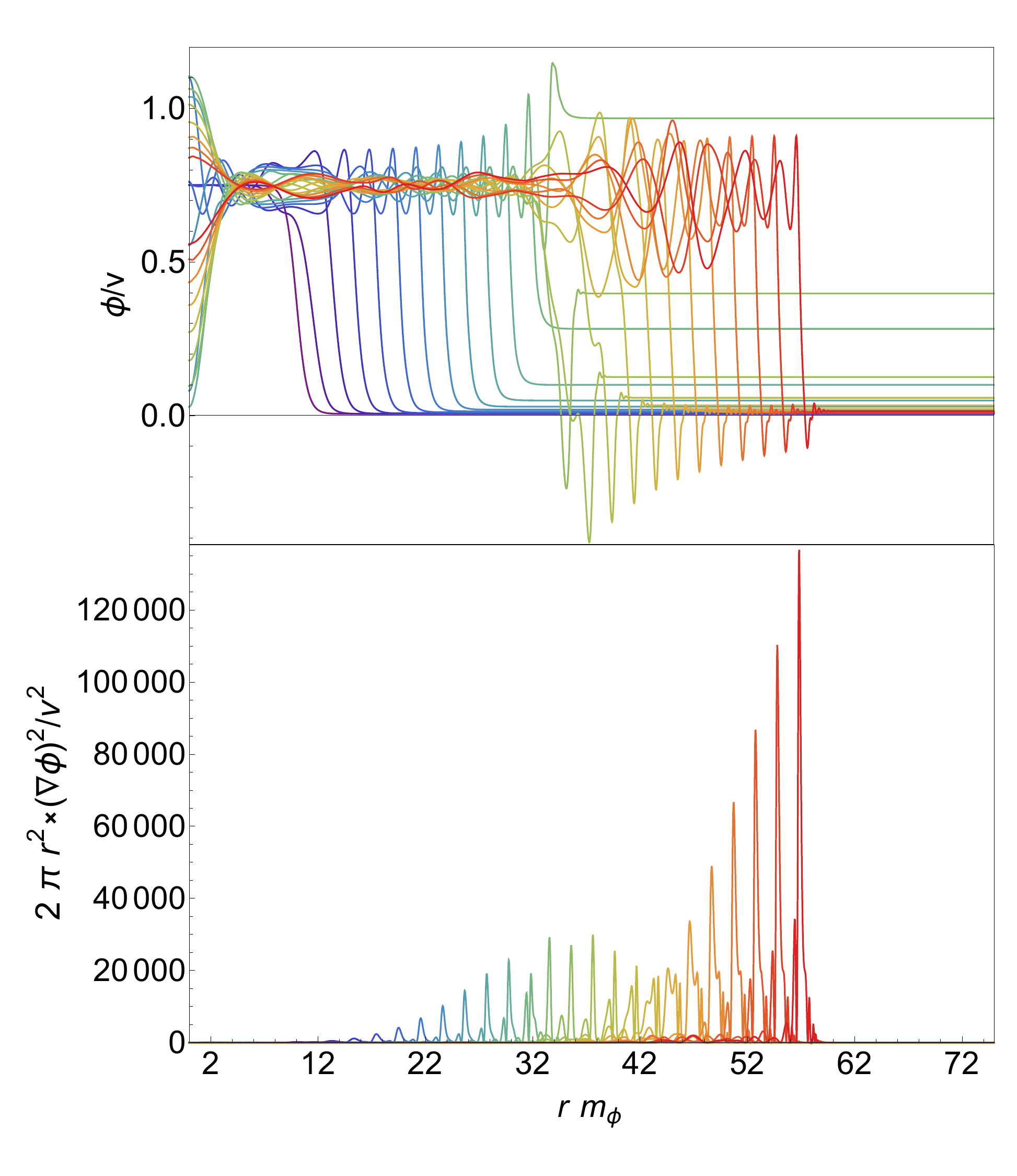}
    \vspace{-10mm}
    \end{center} 
\caption{
    The evolution of a spherically symmetric vacuum bubble. The upper panel shows the field profile $\phi(r)$, while the lower panel displays the gradient energy density on spherical shells, $2\pi r^2 (\nabla\phi)^2$.
    We set the initial field value outside the bubble to $\overline{\phi}_{i,\mathrm{out}} = v/200$ and the initial bubble radius to $r_{i,b} = 10/m_\phi$ with the potential of \Eq{benchmarks} of $n=3/2$. The color changes from purple to red as time progresses. The time slices are sampled as $\delta t = 2/m_\phi$ with $m_\f=(3/2) \sqrt{V_0}/v$. 
}
\label{fig:bubble} 
\end{figure}

This behavior is more clearly verified using one-dimensional simulations of spherically symmetric
wave configurations in the case $n=3/2$, as shown in Fig.~\ref{fig:bubble}, by solving
\beq
\laq{eq1d}
\left(\partial_t^2 - \frac{1}{r^2}\partial_r\!\left(r^2 \partial_r\right) \right)\phi + V_\phi = 0.
\eeq
In this figure, we adopt the potential given on the left-hand side of \eqref{eq:benchmarks}.
One can see that after the onset of oscillations, the shock-wave front continues to propagate as before,
and the gradient energy carried by the wave front continues to increase.

We find that sufficiently deep inside the expanding region, 
the field inside the bubble approaches the vacuum value, 
with only small residual oscillations, 
even after the surrounding field begins to oscillate.\footnote{
An exception to this behavior may arise when the potential possesses a symmetry 
such that the field oscillates into a region where it remains near a point 
with small curvature for an extended period, 
as in ALP hilltop inflation~\cite{Daido:2017wwb,Daido:2017tbr,Takahashi:2019qmh}. 
In this case, domain walls can form as a consequence of nonlinear wave propagation. 
In particular, when a propagating wave encounters an oscillating background outside the bubble, 
the field can be further lifted by the already accelerated wave front. 
As a result, the local field value can exceed the threshold required for domain-wall formation. 
The energy is then stored both in the wave front and in the resulting Baumkuchen domain walls, 
and the associated gravitational-wave signal can be significant~\cite{Narita:2025jeg,Miyazaki:2025tvq}.}

\subsection{Stage 3: Bubble collisions and particle production}

Within a single Hubble patch, there are only a few seeded bubbles that 
expand significantly and absorb most of the vacuum energy. 
After the bubbles expand over a distance of $\mathcal{O}(1/H_{\rm inf})$, 
they inevitably collide with one another. 
This collision process is highly nonlinear and extremely violent. 
Indeed, even before the collision, 
the region with small $\phi$ located between the bubble walls 
can be pushed into the $\phi<0$ region.

In particular, for integer $n$, the locally realized large amplitude due to the bubble gradient makes the scalar field easily settles into another vacuum separated by the hilltop for the inflation.
This causes domain wall formation.\footnote{
This behavior is governed by dynamics different from the hill-climbing phenomenon driven by tachyonic enhancement followed by oscillations~\cite{Antusch:2015vna, Yin:2024pri},
in which fluctuations overshoot the hilltop when the scalar field returns toward it.
}
 Therefore the low scale hilltop inflation generically predicts the sub-Horizon size domain wall formation (see the panels for $t m_\f\geq 370$ in \textcircled{4} in Fig.\ref{fig:lattice2}).
Due to the population bias the domain wall collapse soon, even if the inflaton potential has $Z_2$ symmetry.\footnote{This means if the inflaton is some Higgs field spontaneously breaking a symmetry $\cal G$, topological defects relevant to ${\cal G}\to {\cal H}$ may be induced with $\cal H$ being the unbroken part of the symmetry. A dark monopole dark matter may be obtained if, e.g., ${\cal G}=\SU(2), {\cal H} =\U(1)$ and $\phi$ is the triplet Higgs (see a similar phenomenon for cosmic string \cite{Yin:2024txg,Yin:2024pri}). }

As in the \textcircled{2} and \textcircled{3} in Figs.\ref{fig:lattice} and \ref{fig:lattice2}, one can see that the most energies are eventually stored in the particle kinetic and gradient energies, i.e., relativistic particles are produced, for the energy conservation. Therefore the preheating completes due to the bubble expansion. The remains are energetic particles and waves.

\section{Gravitational waves}

Indeed, our numerical simulations show that the gradient energy,
which constitutes approximately $10\%$ of the total energy density,
is predominantly stored in the bubble walls
(see the comparison between \textcircled{2} and \textcircled{4}
around the bubble collision in Figs.~\ref{fig:lattice} and \ref{fig:lattice2}).
We therefore expect the gravitational-wave production to be significant.

By analogy with a first-order phase transition,
the gravitational-wave abundance can be estimated at the order-of-magnitude level as
\beq
\Omega_{\rm GW}\sim 10^{-2}
\left(\frac{\epsilon_{\rm GW}}{0.1}\right)^2
\left(\Delta t\, H_{\rm inf}\right)^2 ,
\eeq
where $\epsilon_{\rm GW}$ denotes the efficiency of gravitational-wave production. Although a detailed analysis of the gravitational-wave spectrum, which is very important for testing low-scale inflation, is left for future work, we assume the efficiency parameter $\epsilon_{\rm GW}$ to account for systematic uncertainties,
including possible deviations from the standard bubble-expansion picture of ordinary first-order phase transitions.

The typical duration of the source is $\Delta t \sim H_{\rm inf}^{-1}$,
since the separation between bubbles is set by the original instability scale,
which is of order $H_{\rm inf}$ at the end of inflation.
Thus, the characteristic wavenumber of the gravitational waves is expected to be
\beq
k_{\rm GW} \sim H_{\rm inf}.
\eeq

We also note that in integer-$n$ (hilltop) models,
the nonlinear dynamics can lead to the formation of domain-wall configurations
when different vacua are locally populated.
If the vacuum in the $\phi<0$ region corresponds to the true minimum where the cosmological constant is tuned to be zero,
closed domain-wall structures enclosing true-vacuum regions experience outward acceleration
due to the vacuum-energy difference.
In this case, the subsequent wall expansion and collisions can also generate gravitational waves with a broadly similar spectrum.
In particular, when the vacuum-energy difference is as large as ${\cal O}(V_0)$,
the amplitude can become significant and potentially even dominant.

Assuming instantaneous reheating, which is realized in various low-scale inflation models \cite{Daido:2017wwb,Takahashi:2019qmh,Yin:2022fgo},
the present-day gravitational-wave abundance
with the corresponding peak frequency is estimated by
\beq
\Omega_{\rm GW}^{0,\rm peak} \sim 10^{-7} \(\frac{ \e_{\rm GW}}{0.1}\)^2,
\qquad
f_{\rm GW}^{0,\rm peak}\sim
3\times 10^{-6}\,{\rm Hz}\,
\frac{V_0^{1/4}}{240\,\GEV}.
\eeq

\begin{figure}[!t] 
    \begin{center}
        \includegraphics[width=170mm]{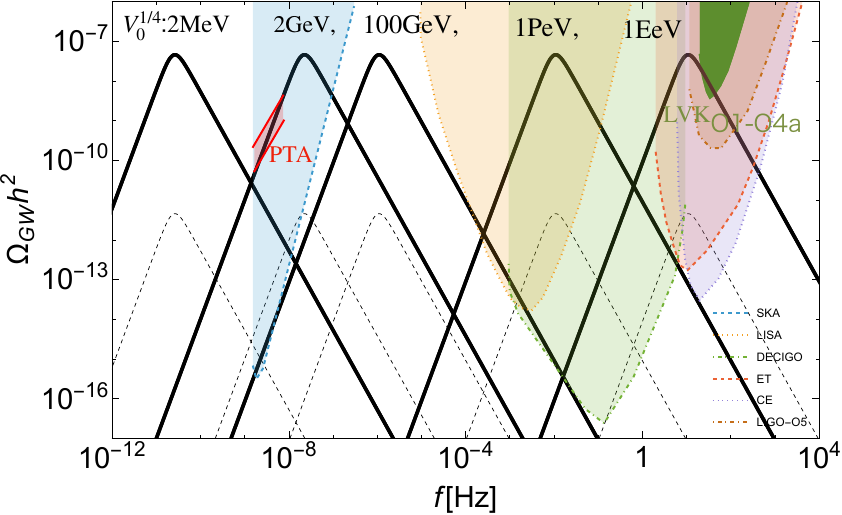}
\vspace{-10mm}
    \end{center} 
\caption{
The gravitational wave prediction from single field low scale inflation by assuming the instantaneous reheating. $\e_{\rm GW}=0.1$ and $\e_{\rm GW}=0.001$ are shown by solid lines and dashed lines, respectively. 
Also shown are limits/preference from the PTA results as well as the limit set by LVK collaboration~\cite{LIGOScientific:2025bgj}, and future reaches of SKA~\cite{Janssen:2014dka,Weltman:2018zrl},  LISA~\cite{LISA:2017pwj}, DECIGO~\cite{Kawamura:2011zz}, ET~\cite{Punturo:2010zz,Sathyaprakash:2012jk}, CE~\cite{Evans:2021gyd}, and LIGO-O5~\cite{TheLIGOScientific:2014jea}. One can see the whole region satisfying \Eq{nonlinear}, can be covered, and the 1EeV is already constrained by the LVK collaboration while 2GeV is favored by the PTA result~\cite{NANOGrav:2023gor,Antoniadis:2023ott,Reardon:2023gzh,Xu:2023wog}. }
    \label{fig:GW} 
\end{figure}

An analytic estimate of the resulting gravitational-wave signal is presented in Fig.~\ref{fig:GW} by taking $\e_{\rm GW}=0.1, 0.001$ for the solid and dashed lines.
Our three-dimensional lattice simulations (see Appendix~\ref{app:1}) for this scenario
exhibit the same large-$k$ scaling, $\propto k^{-2}$,
as found in Ref.~\cite{Konstandin:2017sat} for relativistic bubble walls with $v_{\rm wall}\to 1$,
while modes with wavenumbers smaller than the peak scale as $k^{3}$,
as required by causality.
Also shown are current hints, existing constraints, and future sensitivities
\cite{LIGOScientific:2025bgj,
Janssen:2014dka,Weltman:2018zrl,
Kawamura:2011zz,
Punturo:2010zz,Sathyaprakash:2012jk,
Evans:2021gyd,LISA:2017pwj,
TheLIGOScientific:2014jea,
NANOGrav:2023gor,Antoniadis:2023ott,Reardon:2023gzh,Xu:2023wog}.
Inflationary models with $V_0^{1/4}\lesssim 2\,\MEV$ are excluded,
since they lead to reheating temperatures that are too low to be consistent with
big-bang nucleosynthesis. 

Our results highlight that low-scale inflation, despite the absence of observable primordial tensor modes in the CMB, can—at least in certain situations—possess distinctive and testable predictions in the gravitational-wave sector.
This opens a new observational window for probing inflation at MeV--EeV energy scales.
Such signals can be searched for by current and future gravitational-wave experiments, including ongoing searches by the LIGO--KAGRA--Virgo collaboration~\cite{LIGOScientific:2025bgj}, pulsar timing arrays, and future detectors such as SKA~\cite{Janssen:2014dka,Weltman:2018zrl}, LISA~\cite{LISA:2017pwj}, DECIGO~\cite{Kawamura:2011zz}, ET~\cite{Punturo:2010zz,Sathyaprakash:2012jk}, and CE~\cite{Evans:2021gyd}.

In fact, existing LIGO--KAGRA--Virgo data already constrain some inflationary models with
$V_0^{1/4}\approx 10^9\GEV$,
which lies close to the interesting region where QCD axion dark matter may originate from a string axion~\cite{Graham:2018jyp,Takahashi:2018tdu}.
We note that for $n=3/2$ this limit does not apply (see \Eq{nonlinear}), whereas for $n\geq 2$ it does.
On the other hand, the nanohertz stochastic gravitational-wave background detected by pulsar timing arrays can be explained by inflation at the ${\cal O}(\mathrm{GeV})$ scale, as suggested in scenarios motivated by the trans-Planckian censorship conjecture~\cite{Bedroya:2019snp,Bedroya:2019tba,Marsh:2019bjr}.

Low-scale inflation generically predicts a relatively large running of the scalar spectral index. If precise theoretical predictions of the gravitational-wave signal can be established and matched by observational detection, the combination would provide a robust and distinctive signature of single-field low-scale inflation.

\section{Conclusions and Discussions}
\lac{discussions}
We have shown that a broad class of low-scale single-field inflation models exhibits an extremely efficient tachyonic instability at the end of inflation. Once the inflationary energy scale is sufficiently low, the tachyonic growth of inflaton fluctuations becomes nonlinear before the onset of any coherent homogeneous oscillation. As a result, the post-inflationary dynamics is dominated by strongly inhomogeneous configurations, leading to the formation of relativistically expanding bubble-like structures and nonlinear shock waves. We referred to this phenomenon as a tachyonic shock.

A key feature of this dynamics is that it is governed by the Hubble scale rather than by the microscopic mass scale of the inflaton. Consequently, the characteristic length and time scales of the nonlinear evolution, as well as the peak frequency of the resulting gravitational waves, are set by the inflationary Hubble parameter. This is in sharp contrast to conventional tachyonic or parametric preheating scenarios, such as those realized in hybrid inflation, where the dynamics is controlled by the mass scale of the waterfall field and leads to gravitational waves at higher frequencies.

The bubble-wall--like structures formed during the tachyonic shock efficiently convert the vacuum energy into gradient energy and subsequently into gravitational waves, in close analogy with a strongly first-order phase transition but without tunneling or barrier crossing. Remarkably, this mechanism allows inflationary models at MeV--EeV energy scales to be probed by gravitational-wave observations, despite the absence of observable primordial tensor modes in the CMB. Current limits from the LIGO--KAGRA--Virgo collaboration already constrain this parameter space, while the stochastic gravitational-wave background reported by pulsar timing arrays may be interpreted as a signal from inflation at the GeV scale.


\paragraph{More General Potentials}
So far, we have illustrated the tachyonic shock phenomenon by focusing on inflationary potentials
dominated by a single term of the form $\phi^{2n}$ in the relevant field range.
More generally, however, a realistic inflationary potential can be written as a Taylor expansion,
\beq
V(\phi)=V_0-\sum_{n} \lambda_{(n)} \phi^{2n},
\eeq
where multiple terms may be required in order to reproduce the spectral index. Again $n$ takes half integer or integer. 
Here, we show that, unless fine-tuned cancellations are imposed, the qualitative behavior found in the single-term analysis
persists for such general potentials.

Suppose that the amplitude of the curvature perturbation is dominantly determined 
by a single term $\phi^{2n_0}$ around horizon exit.
Then the corresponding coupling $\lambda_{(n_0)}$ is fixed by the CMB normalization
through \Eq{lambda} for a given vacuum energy $V_0$.
Since $\phi$ increases with time and the terms with $n<n_0$ are subdominant at horizon exit,
they remain subdominant at larger field values.

For terms with $n>n_0$, the corresponding couplings $\lambda_{(n)}$ are generically smaller
than the values that would be obtained from \Eq{lambda} evaluated for the same $n$,
unless additional tuning is introduced.
This is because the field value at horizon exit does not vary significantly
for \Eq{lambda} case by changing $n$ once $V_0, n_s$ and $A_S$ are fixed,
and therefore $V'$ and $V''$ are essentially determined by $V_0$ and $A_S$. Then the field value $\sim V'/V''$ does not change much by varying $n$ for the single term case. 
Consequently, in order to neglect the contribution of higher-order terms,
their couplings $\lambda_{(n)}$ are small or comparable to the \Eq{lambda} case.

As long as the slow-roll evolution proceeds toward $\phi>0$,
terms with $n<n_0$ are irrelevant, while terms with $n>n_0$ may become important at later stages of the rolling.
These higher-order terms may control the end of inflation and modify the field excursion after inflation.
If different terms dominate successively during the rolling phase, the tachyonic instability amplifies
field fluctuations in a multiplicative manner.
For example, if the dominant contribution for the increasing amplitude of the tachyonic mass changes as
$\phi^{2n_0}\to \phi^{2n_0+1}\to \phi^{2n_0+2}\to\cdots\to \phi^{2n_{\rm end}}$,
the total enhancement of fluctuations is given by the product of the growth factors in each interval.

Although the linear growth scales as $\delta\phi \propto \bar\phi^n$ and does not depend on $\lambda_{(n)}$ explicitly,
$\lambda_{(n)}$ affects the amplification indirectly through the background evolution $\bar\phi(t)$.
In particular, for smaller $\lambda_{(n)}$ with $n>n_0$, that becomes relevant at later time,  the field excursion in the regime becomes longer, and hence enhances the total growth factor.

Since higher powers of $\phi$ in the single-term case lead to stronger tachyonic amplification for fixed $V_0$,
as shown in Fig.~\ref{fig:1},
the total enhancement in the general case is larger than that obtained in the single-term model,
$\left(\bar\phi_{\rm end}/\bar\phi_{\rm inf}\right)^{n_0}$.
Therefore, the single-term analysis provides a conservative estimate,
and the tachyonic shock phenomenon is a generic outcome for an even broader class of low-scale single-field
inflationary potentials than considered in the main text.

\paragraph{New phenomena related to tachyonic shocks}
Let us mention several related aspects of the tachyonic shock phenomenon that we have already investigated or plan to study in future work.
\begin{itemize}
\item In the case of ALP hilltop inflation~\cite{Daido:2017wwb,Daido:2017tbr,Takahashi:2019qmh}, which possesses multiple vacua due to its periodic structure, Baumkuchen domain walls~\cite{Narita:2025jeg,Miyazaki:2025tvq} are found to form as a consequence of the asymmetrically enhanced field amplitudes.

\item The dynamics described above generates an ultra-relativistic bubble expansion. As a consequence, a wide range of phenomena can potentially occur, including dark matter production~\cite{Falkowski:2012fb, Azatov:2021ifm,Baldes:2022oev,Azatov:2022tii,Azatov:2023xem,Azatov:2024crd,Ai:2024ikj}
and baryogenesis~\cite{Azatov:2021irb,Baldes:2021vyz,Azatov:2023xem,Chun:2023ezg}.
\item Our main finding---that tachyonic instability generically leads to a strong enhancement of small-scale inhomogeneities in low-scale inflation---
implies that thermalization may proceed in a highly asymmetric manner
(see, e.g., Ref.~\cite{Kurkela:2011ti}), in sharp contrast to the usual homogeneous assumptions. Even after the inflaton reaches to the homogeneous regime, the momentum of fragmented inflaton is boosted and the reheating rate must be different as usual assumption. 
\item This phenomenon can also have important implications for other cosmological processes, such as baryogenesis and primordial black hole formation.
For instance, if the inflaton couples to the Higgs field, inflaton-nonlinear shock waves can locally restore the electroweak symmetry by generating induced Higgs bubble walls (see~\cite{Lee:2024xjb}).
This may allow sphaleron processes to remain efficient even when the inflation scale—and hence the reheating temperature—is below the weak scale.  Moreover, in the case of integer $n$ where the vacuum energies on the two sides of the hilltop are different, we have checked that genuine bubble expansion can occur after domain-wall formation.
Such dynamics may generate overdensities in certain Hubble patches, potentially leading to primordial black hole formation.
\end{itemize}

Overall, our results indicate that low-scale inflation can exhibit nontrivial and potentially testable post-inflationary dynamics that have not been fully explored.
In particular, tachyonic shocks may provide a generic mechanism that connects low-scale inflationary physics with present-day gravitational-wave observations.
If confirmed by more detailed numerical studies, this mechanism could offer a new way to probe the physics of the very early Universe.

\section*{Acknowledgement}
This work is supported by JSPS KAKENHI Grant Nos. 22K14029 (W.Y.), 22H01215 (W.Y.), Graduate Program on Physics for the Universe (Y.N.), and JST SPRING, Grant Number JPMJSP2114 (Y.N.). W.Y. is also supported by Selective Research Fund from Tokyo Metropolitan University.

\appendix
\section{Setups of Numerical simulation}

\label{app:1}
To show this happen, we use the classical lattice simulation by modifying {\tt CosmoLattice}~\cite{Figueroa:2020rrl,Figueroa:2021yhd}. 
We note that there are very large hierarchy in the system. We need the vacuum mass $m_\phi\gg H_{\rm inf}$, and the Lorentz factor at the collision is $\gamma_w \sim m_{\phi}/H_{\rm inf}$, and thus the shock wave width is $(m^2_{\phi}/H_{\rm inf})^{-1}$. More crucially, to have our mechanism work we need $m_\f\gg 10^6 H_{\rm inf}$ for $n=3/2~(2)$ by requiring  $V_0^{1/4}\ll 10^5 (10^{9})\GEV$. 
Therefore a realistic lattice needs $N\gg ((10^6)^2)^3$ points, which is not possible for us. 

Thus, we use the following trick to induce the same phenomenon in a toy setup. In this part, we consider the period $\bar{\phi}> \phi_{\rm inf}$. 
Since the rolling time scale is much shorter than the expansion rate, we do not consider the expansion of the Universe.

The equation of motion is independent of the overall scale $V_0$ once we write
$V=V_0\,F(\phi/v)$, where $F$ is a dimensionless function of $\phi/v$.
The dependence on $V_0$ and $v$ can be absorbed into a rescaling of time and space coordinates.
In terms of the dimensionless field $\hat{\phi}\equiv \phi/v$,
the equation reduces to
$\hat\Box \hat{\phi} = \partial_{\hat{\phi}} F(\hat{\phi})$,
where $\hat\Box$ denotes the d'Alembertian operator with coordinates rescaled by $V_0/v^2$.
When discussing gravitational waves, however, the absolute energy scale will become relevant.

\begin{itemize} 
\item  To get enough fluctuation for the tachyonic shock to happen, we enhance the initial fluctuation by hand, $
 \delta \phi_k = \frac{1}{\sqrt{k}} c \times \tl R
$
where $\tl R$ is gaussian Random noise which is typically $1$. This resembles the Minkowski fluctuation with $c=1$ but we include $c>1$ to enhance the fluctuation, so that the bubble formation can happen within the simulation time. This is taken to be $3\times 10^6 (4\times 10^6)$ for $n = 3/2 (2)$, while the initial value of $\hat{\bar\phi}=0.004 (0.006).$ Taking $c=0$ fixing other parameters, the $\bar \f$ evolution is verified to happen at much later time in both simulations.

\item We also set an UV cutoffs, $k_{\rm UV}$ for $k$ to avoid unphysical artifacts and to avoid the renormalization in the classical lattice simulation \cite{Felder:2000hj,Rajantie:2000nj,Berges:2013lsa}. 
We take $k_{\rm UV} = 0.07 (0.05)m_0$ for $n=3/2 (2)$ which is close to $m_{\rm eff}$ at the moment $\bar \phi\sim \sigma_\phi$. The lowest scale that is enhanced, which is roughly, $0.05 m_0$ estimated from $m_{\rm eff}$ by the initial condition for $n=3/2$. In the $n=2$ case the IR is cutoff by the box size $L$ satisfying $2\pi/L=0.01 m_0$, which is also the box size for $n=3/2$. 
This lets the bubble to get a Lorentz factor of $10-100$.

\item We take the simulation in 2D spatial space and $N=8096^2$ so that we are allowed to have relatively good resolution to even cover the bubble wall taking account of $10-100$ Lorentz contract. 

\item 
We have checked that the bubble expansion remains intact when we slightly vary $k_{\rm UV}$.
In particular, choosing $k_{\rm UV}\sim m_{\rm eff}$ at the end of Stage~1 is not very reasonable,
since modes near the cutoff are not sufficiently tachyonically enhanced. 
Fixing the IR cutoff and decreasing $k_{\rm UV}$ does not change the bubble-expansion dynamics. 

\item  We have also verified that adopting the spectrum derived from the linear analysis of Stage~1,
\beq
\delta\phi_k \propto k^{-\frac{n}{n-1}-\frac{1}{2}},
\eeq
does not qualitatively modify our conclusions; if anything, it makes the Hubble-scale bubble expansion more robust,
because the spectrum suppresses small-scale fluctuations and hence reduces the formation of smaller bubbles. The animations in the supplementary material are made with this initial condition. For the $n=5$ case (the file \texttt{n5.gif}), we use the potential $F= (3125/46656-\hat{\f}^{10}+\hat{\f}^{12})$.
\item  For some cases with $n>2$, as well as in ALP hilltop inflation models with
$F= \cos(\hat\phi) - \cos(2\hat\phi)/4 + 5/4$,
we also observe similar dynamics (and the Baumkuchen domain-wall formation for the ALP hilltop inflation).
Bubble formation in a runaway potential,
$F= (1 - \tanh[n \hat\phi])$,
has been checked as well. These results, together with further details, will be presented in future work.
\end{itemize}

\begin{figure}[!t] 
    \begin{center}
        \includegraphics[width=78mm]{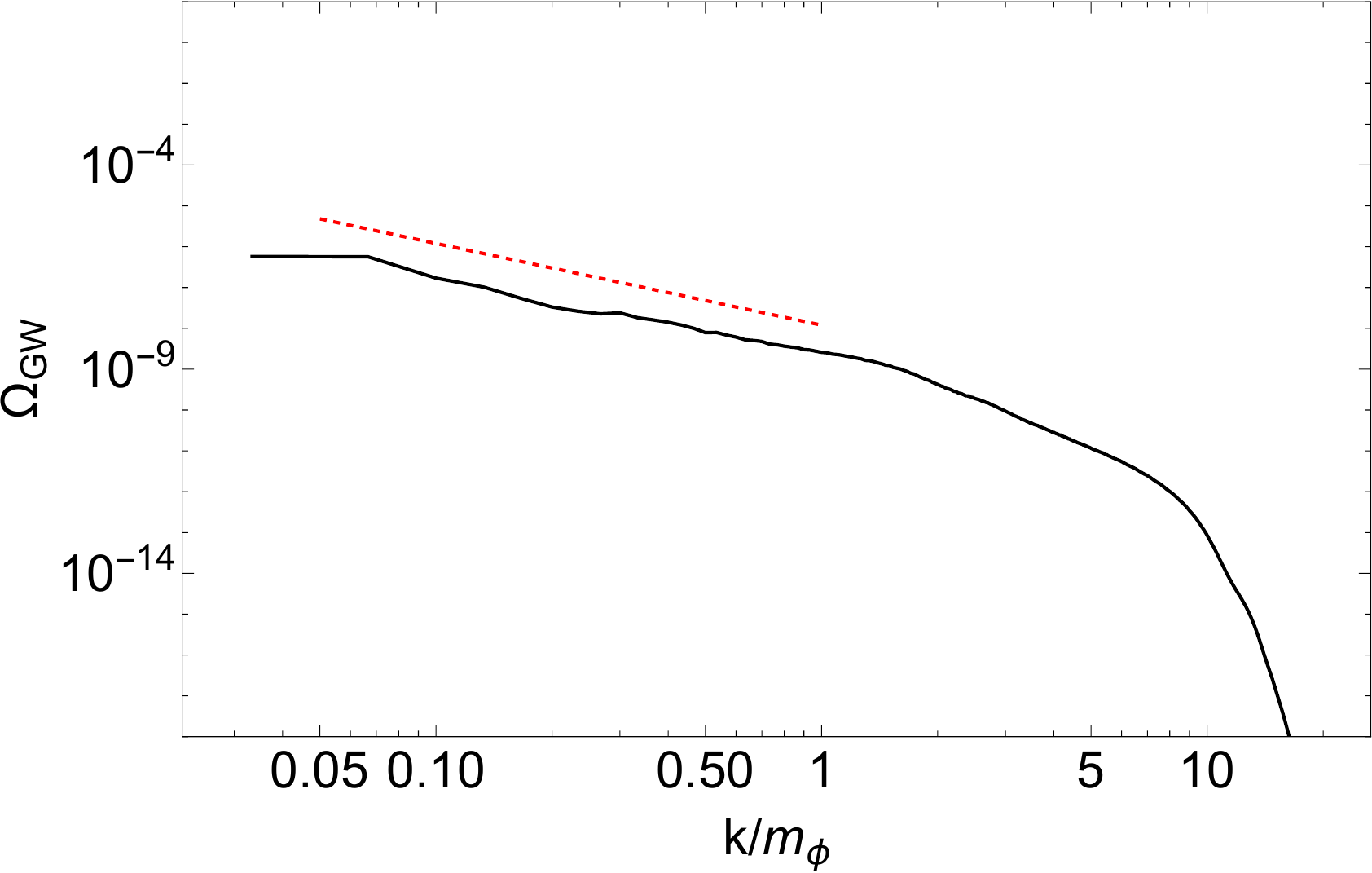}
        \includegraphics[width=75mm]{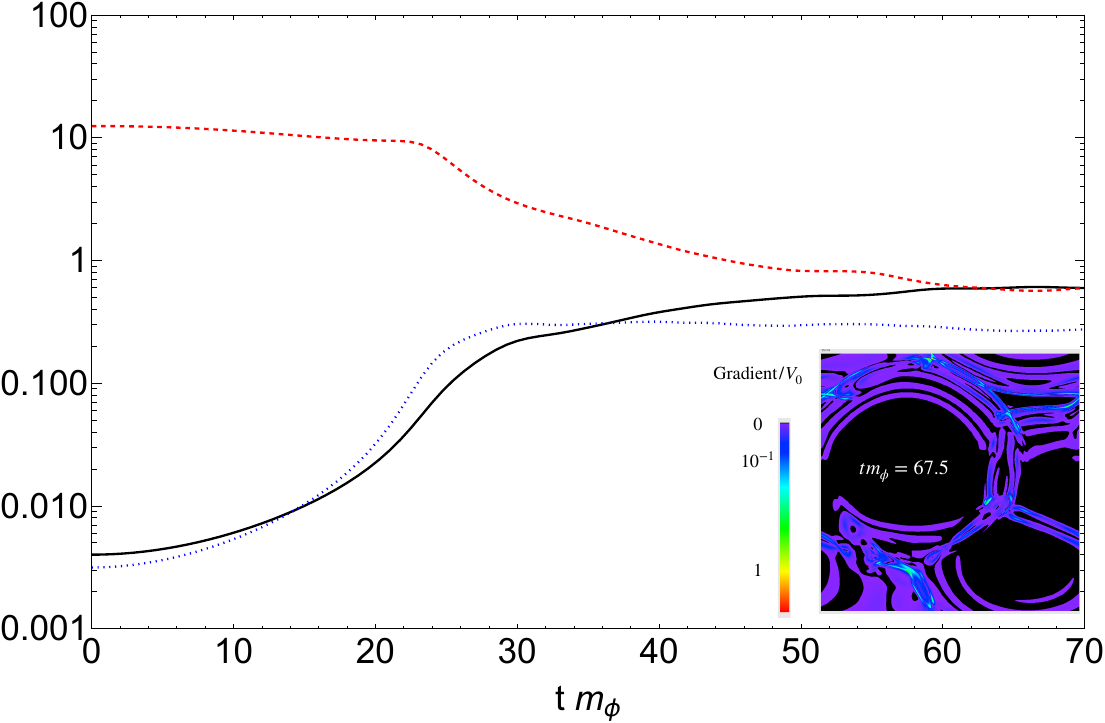}
\vspace{-5mm}
    \end{center} 
\caption{
The gravitational wave from the 3D numerical simulation with $n=3/2$ (black solid line in the left panel). Also shown is the $k^{-2}$ line in red dashed line. We  note that there are $\O(10\%)$ uncertainty in the power. Left panel shows time evolution of the scalar field, and the corresponding snapshot at $t m_\phi=67.5$. Here $m_\phi=3/2m_0=3/2 \sqrt{V_0/v^2}$.}
    \label{fig:GW2} 
\end{figure}

The same procedure is used to estimate the high-frequency regime of the gravitational-wave spectrum
in a three-dimensional $800^3$ simulation with $n=3/2$.
We take $2\pi/L = 0.05\, m_0$, $k_{\rm UV} = 0.1\, m_0$, $\bar{\phi}_i = 0.004$, \AND\ $c = 3\times 10^7$.
To compute the gravitational-wave signal, the Planck scale must be specified explicitly,
so the previously irrelevant parameters $v$ and $V_0$ become relevant.
Here we choose $v = 10^{16}\,\GEV$ \AND\ $V_0 = 10^{48}\,\GEV^4$.
The result is shown in Fig.~\ref{fig:GW2}, together with the time evolution of the scalar field (right panel).

\bibliography{GenericALPDMbound.bib}
\end{document}